\documentclass[aps,prd,onecolumn,groupedaddress,showpacs,nofootinbib,amssymb]{revtex4}
\usepackage{amsmath}
\usepackage{appendix}
\usepackage{multirow}
\usepackage{amsfonts}
\usepackage{xcolor}
\usepackage{comment}
\usepackage{graphicx}
\usepackage{siunitx}
\usepackage{lineno}
\usepackage{subfigure}
\usepackage{float}
\usepackage{ulem}

\usepackage[
    pdfauthor={Teng Zhang},
    pdftitle={Acc},
    colorlinks=true,
    linkcolor=black,
    urlcolor=blue,
    citecolor=blue
]{hyperref}

\begin{document}

\title{Study of acceleration measurement in gravitational wave detection}

\author{Junlang Li$^1$ \& Teng Zhang$^2$}
\address{$^1$Gravitational Wave and Cosmology Laboratory, Department of Astronomy, Beijing Normal University, Beijing 100875, China}

\address{ $^2$School of Physics and Astronomy, and Institute for Gravitational Wave Astronomy, University of Birmingham, Edgbaston, Birmingham B15\,2TT, United Kingdom}
\tolerance=5000
\begin{abstract}
Position-meter and speed-meter interferometers have been analysed for detecting gravitational waves. We introduce the concept of acceleration measurement in comparison with position and speed measurement. In this paper, we describe a general acceleration measurement and derive its standard quantum limit. We provide an example of an acceleration-meter interferometer configuration. We show that shot noise dominates at low frequency following a frequency dependence of $1/\Omega^2$, while radiation pressure noise is constant. The acceleration-meter has even a stronger radiation pressure noise suppression than speed-meter. 
\end{abstract}
\vspace{2pc}

\maketitle

\section{Introduction}

Quantum noise has been one of the most important noise sources over the whole frequency band in laser-interferometric gravitational wave detectors. It contains two parts: shot noise (sensing noise), and radiation pressure noise (back-action noise). Shot noise comes from the fluctuation of the photon number of a laser. It is proportional to the square root of the laser power, $1/\sqrt{P}$, and dominant in high frequency band. Radiation pressure noise is induced by the back action of the laser beam onto the test mass. The radiation pressure noise is proportional to $\sqrt{P}$ and is dominant in the low frequency band. At each frequency, there exists an optimal power, which determines the minimal sum of shot noise and radiation pressure noise. This forms the so called Standard Quantum Limit (SQL) as a consequence of the Heisenberg uncertainty principle\,\cite{braginsky_khalili_thorne_1992}.

The SQL can be overcome by conducting a so called quantum nondemolition measurement\,\cite{Braginsky547}, in which the uncertainty of the observable is beyond the SQL, \textit{e.g.}, by measuring the momentum of a free test mass. Speed measurement is actually a good substitution that is approximately proportional to momentum\,\cite{BRAGINSKY1990251,2018LSA.....7...11D}. Until now there are already several types of proposed speed-meter, such as the sloshing speed-meter\,\cite{1999Dual,Freise_2019,PhysRevD.66.122004,PhysRevD.86.062001}, Sagnac interferometer\,\cite{2002Sagnac}, polarisation circulation speed-meter\,\cite{2018LSA.....7...11D},  and EPR-type speed meters\,\cite{KNYAZEV20182219}. The relevant study in theory and experiment has also been continued in recent years\,\cite{PhysRevD.89.062009,Gr_f_2014,Danilishin2019,Huttner_2016,Zhang_2018}. The signal and radiation pressure noise of gravitational wave detectors comes from the response of the detector to the gravitational wave tidal force and laser radiation pressure force induced mirror displacement. In the speed meter, the radiation pressure force is partially cancelled by adopting speed as the observable, which leads to improvement of the signal to noise ratio in the frequency band where radiation pressure noise dominates. It is instructive to consider what can be achieved by acceleration measurement. 

In this paper, we develop the concept of the acceleration measurement and show a realisation of an acceleration-meter evolving from the Sagnac speed-meter and polarisation circulation speed-meter, called the " polarisation-Sagnac acceleration-meter". In Sec.~\ref{sec:Conceptual scheme}, we first study the acceleration measurement as a supplement to describe the dynamics of a linear system in addition to position and speed measurement. In Sec.~\ref{sec:Design of acceleration-meter}, we describe the design of the polarisation-Sagnac acceleration-meter interferometer and calculate the quantum noise by deriving the Input/Output (I/O) relations.

\begin{figure}[ht]
\centering
\includegraphics[width=\columnwidth]{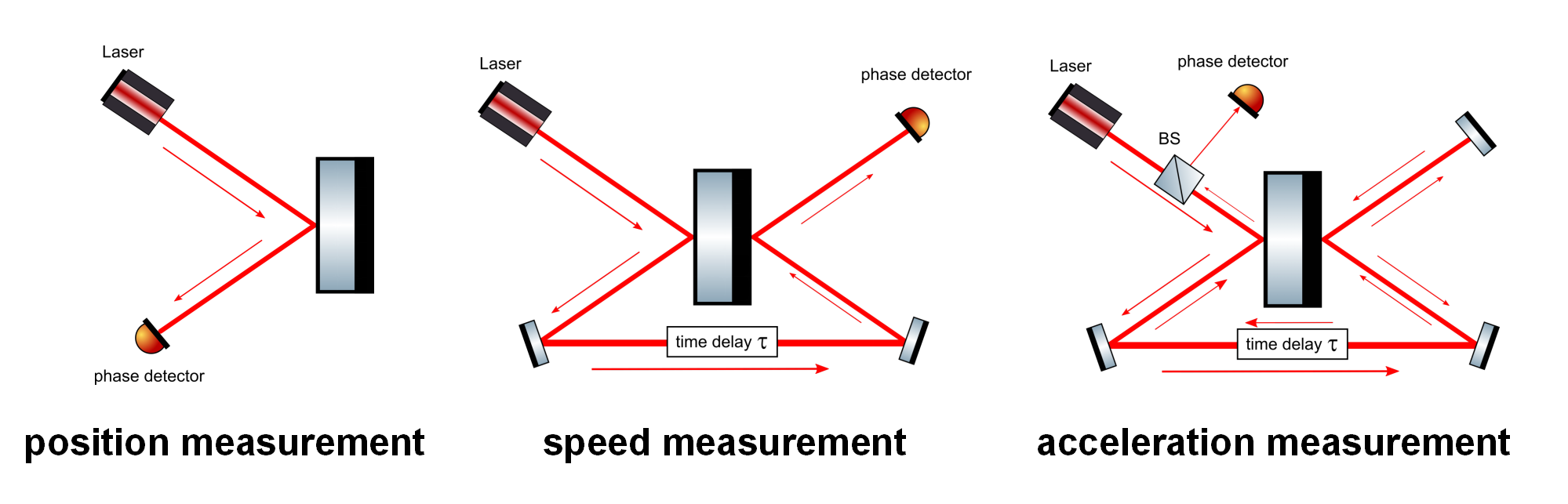}
\caption{Conceptual scheme of position, speed and acceleration measurements. In position measurement, light reflects once on one side of the mirror. In speed measurement, light reflects on two sides of the mirror with time delay $\tau$. In acceleration measurement, light reflects twice on each side of the mirror.}
\label{fig:accelerationmeter}
\end{figure}

\section{Conceptual scheme of acceleration-meter}\label{sec:Conceptual scheme}
We start from a simplified position measurement scheme that includes all relevant key features. As shown in left picture in Fig.~\ref{fig:accelerationmeter}, a laser beam is detected after reflecting from the test mass. The displacement of test mass can be measured with a phase detector. The power spectral density of the total noise for a position measurement can be written as\,\cite{2012LRR....15....5D}:
\begin{equation}
S_{\rm{pos}}(\Omega)=S_{x}+|\chi(\Omega)|^{2}S_{F}+2\mathbf{Re}[\chi^*(\Omega)S_{xF}]\,,
\label{sd}
\end{equation}
where $S_{x}$ is the shot noise spectral density in displacement, $|\chi(\Omega)|^{2}S_{F}$ is the back action noise spectral density, $S_{xF}$ is the cross-correlation noise spectral density. $\chi$ is the susceptibility that quantified the response of displacement to force, which is $-1/(M\Omega^{2})$ for a free mass. The appearance of  $S_{xF}$ relates to the readout quadrature of the laser field. For a phase quadrature measurement, $S_{xF}=0$, and the spectral density reads (see \ref{Position measurement} for detailed derivation),

\begin{equation}
S_{\rm{pos}}(\Omega)=S_{x}+|\chi(\Omega)|^{2}S_{F}\geq\frac{\hbar}{M\Omega^{2}}\,.
\end{equation}
At alternative readout quadratures, the resulting measurement uncertainty can beat SQL.

The toy model of a speed measurement is shown in the middle part of Fig.~\ref{fig:accelerationmeter}. The light gets reflected from the front and back of test mass at two moments. It records the speed information of the test mass: $v(t)=[x(t+\tau)-x(t)]/\tau$, where $\tau$ is the time delay between two sensing moments. Similar to the position-meter, the spectral density of the speed-meter can be written in the same form as Eq\,(\ref{sd}):
\begin{equation}\label{SSspe}
S_{\rm{spe}}(\Omega)=\frac{S_{x}}{\Omega^{2}\tau^2}+\Omega^2\tau^{2}|\chi(\Omega)|^{2}S_{F}-2\mathbf{Re}[\chi^*(\Omega)S_{xF}]\,,
\end{equation}
When $S_{xF}=0$, the speed measurement also has the uncertainty relation,
\begin{equation}
S_{\rm{spe}}(\Omega)\geq \frac{S_{x}}{\Omega^{2}\tau^2}+\Omega^2\tau^{2}|\chi(\Omega)|^2\frac{\hbar^{2}}{4S_{x}}\geq\frac{\hbar}{M\Omega^{2}}\,.
\end{equation}

The acceleration measurement can be achieved as shown in right picture in Fig.~\ref{fig:accelerationmeter}. The light first gets reflected on both side of test mass. Then it travels back along the way it comes and get detected by a phase meter. The light field acquires the acceleration information of the test mass: $a(t)=[v(t+\tau)-v(t)]/\tau$. The spectral density can be derived as below(see \ref{Acceleration measurement} for more details),
\begin{equation}\label{SSacc}
S_{\rm{acc}}(\Omega)=\frac{S_{x}}{\Omega^{4}\tau^4}+\Omega^4\tau^{4}|\chi(\Omega)|^{2}S_{F}-2\mathbf{Re}[\chi^*(\Omega)S_{xF}]\,.
\end{equation}
When $S_{xF}=0$, the acceleration measurement is also constrained by SQL,
\begin{equation}
S_{\rm{acc}}(\Omega)\geq \frac{S_{x}}{\Omega^{4}\tau^4}+\Omega^4\tau^{4}|\chi(\Omega)|^2\frac{\hbar^{2}}{4S_{x}}\geq\frac{\hbar}{M\Omega^{2}}\,.
\end{equation}

\section{Polarisation-Saganac acceleration-meter}\label{sec:Design of acceleration-meter}
\begin{figure}[h!]
    \centering
    \includegraphics[width=23pc]{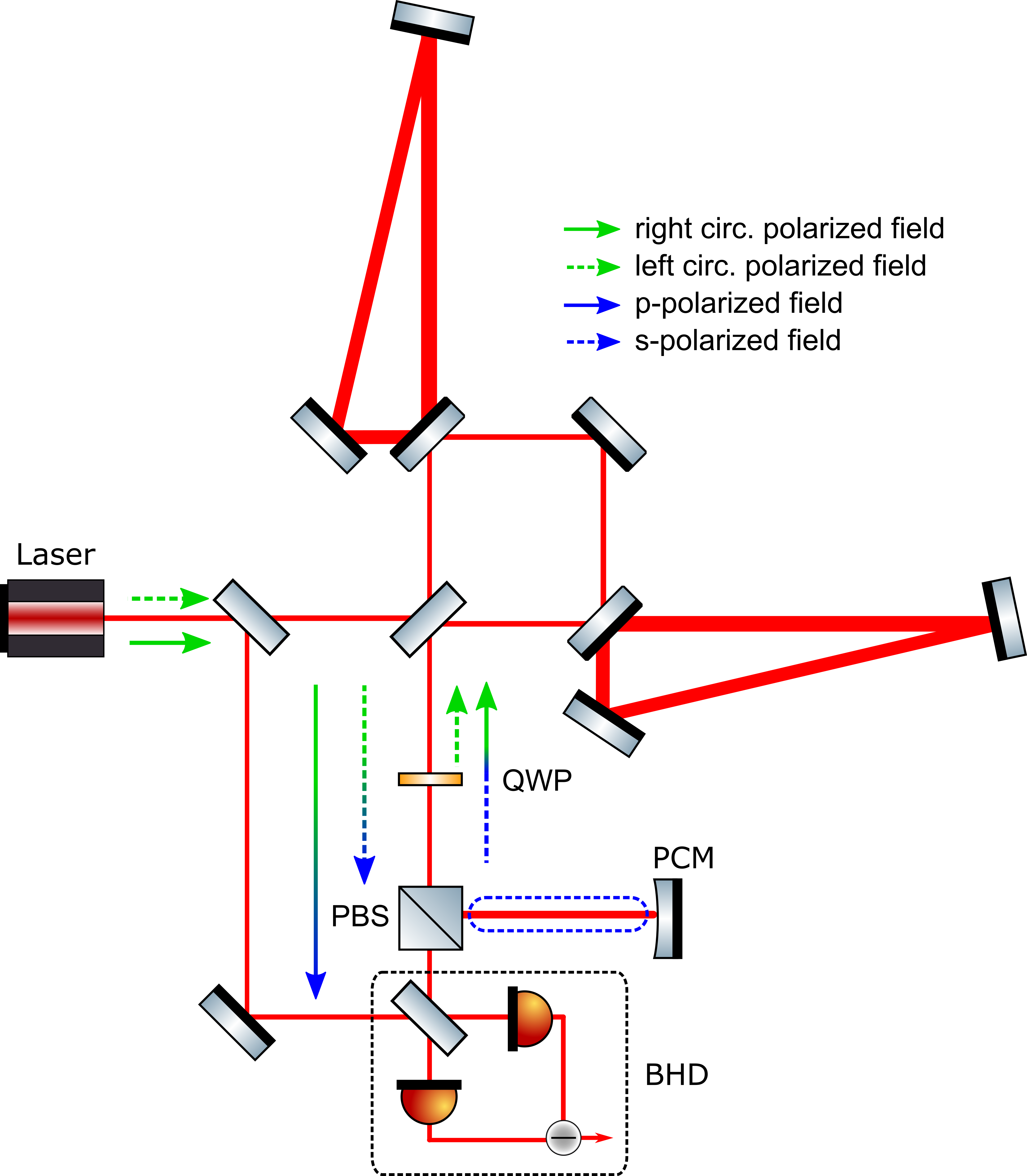}
    \caption{Schematic of acceleration-meter. The acceleration-meter is based on the polarization circulation interferometer and Sagnac interferometer. The red solid line is the carrier light beam, its polarization is shown by the arrows. QWP stands for quarter-wave plate. PBS stands for polarization beam splitter. PCM is a polarization circulation mirror. BHD stands for balanced homodyne detector.}
    \label{fig:02}
\end{figure}

In this section, we describe a realisation of the acceleration-meter interferometer and analyse the quantum noise limited sensitivity by deriving the I/O-relation in the \textit{two-photon formalism}\,\cite{1985PhRvA..31.3068C}.
\subsection{General description}
The acceleration-meter interferometer is based on the polarization circulation interferometer and Sagnac interferometer as shown in Fig.~\ref{fig:02}. In this subsection, we illustrate how the light travels within the detector and conducts an acceleration measurement.

Firstly, the $p$-polarized light emitting from the laser can be regarded as a linear combination of two circularly polarized light, left($l$)-polarized and right($r$)-polarized. Each polarization has only half of the laser power. The $l$-polarized light passes through Sagnac  interferometer and carries velocity information of the differential motion of the arm mirrors $\dot{x}_{\rm dARM}(t)$. At the output of Sagnac interferometer, the $l$-polarized light is transformed into $s$-polarized light by a quarter-wave plate(QWP). It then gets reflected by polarization beam splitter(PBS) and gains a phase shift by reflecting off from the polarization circulation mirror(PCM). After getting transformed into $r$-polarized light by the QWP, it enters the Sagnac interferometer again from the dark port and senses the speed $\dot{x}_{\rm dARM}(t+\tau)$. 
It interferes with the $r$-polarized light that entered the Sagnac interferometer before and comes out as $p$-polarized light. When the phase shift of the sidebands from the PCM equals $2n\pi$ (n is integer), we measure the acceleration $a(t)\propto[\dot{x}_{\rm dARM}(t+\tau)-\dot{x}_{\rm dARM}(t)]$. The minus sign comes from the main beam splitter.

\subsection{Derivation of I/O-relations}

We derive the I/O-relation in the two-photon formalism\,\cite{kimble2001, 1985PhRvA..31.3068C}. The light field can be represented as a combination of a classical mean amplitude and small but non-zero quantum fluctuations. In the two-photon formalism, we can describe the fluctuations by a 2-dimensional vector $\mathbf{a}=\{\hat{a}_{c}, \hat{a}_{s}\}^{T}$, where $\hat{a}_{c,s}$ are the quadrature amplitudes,
\begin{equation}
\begin{aligned}
\hat{a}_{c}=\frac{\hat{a}_{+\Omega}+\hat{a}_{-\Omega}^{\dagger}}{\sqrt{2}}\,,\\
\hat{a}_{s}=\frac{\hat{a}_{+\Omega}-\hat{a}_{-\Omega}^{\dagger}}{i\sqrt{2}}\,
\end{aligned}
\end{equation}
where $\hat{a}$ and $\hat{a}^{\dagger}$ are the creation and annihilation operators, the subscript $\pm\Omega$ represent positive and negative sidebands with respect to the carrier field, and $\Omega$ is the sideband frequency.
A general form of the I/O-relation for an interferometer can be written as,
\begin{equation}
\hat{\mathbf{o}}=\mathbb{T}\hat{\mathbf{i}}+\hat{\mathbf{R}}\frac{x}{x_{\rm SQL}},
\end{equation}
where $\hat{\mathbf{i}}$ and $\hat{\mathbf{o}}$ are the input and output field at the dark port, $\mathbb{T}$ is a 2×2-matrix of the optomechanical transfer function, $\hat{\mathbf{R}}$ is a 2-dimensional vector of the response to $x$, $x$ is the differential mode of displacements of the test masses, and $x_{\rm SQL}=\sqrt{\frac{2\hbar}{M\Omega^{2}}}$ is the standard quantum limit of a free test mass with effective mass $M$.

The I/O relationship of the acceleration-meter scheme can be obtained by applying the Sagnac interferometer I/O-relations for each polarization mode. In the following derivation, we need to remember that both circular polarizations contribute to the common back-action force while only $p-$polarized light enters to the balanced homodyne detection (BHD). The input light field can be represented as a combination of $l-$ and $r-$polarized light
$\hat{\mathbf{i}}=(\hat{\mathbf{i}}_{l}+\hat{\mathbf{i}}_{r})/\sqrt{2}$. Now we can write down the whole I/O-relations:
\begin{equation}\label{IOrelation}
\begin{aligned}
\hat{\mathbf{o}}_{l}&=\mathbb{T}_{\rm{sag}}^{l}\hat{\mathbf{i}}_{l}+\mathbb{T}_{\rm{sag}}^{b.a.}\hat{\mathbf{i}}_{r}+\hat{\mathbf{R}}_{l}\frac{x}{x_{SQL}}\,,\\
\hat{\mathbf{o}}_{r}&=\mathbb{T}_{\rm{sag}}^{b.a.}\hat{\mathbf{i}}_{l}+\mathbb{T}_{\rm{sag}}^{r}\hat{\mathbf{i}}_{r}+\hat{\mathbf{R}}_{r}\frac{x}{x_{SQL}}\,,\\
\hat{\mathbf{i}}_{r}&=\mathbb{P}\hat{\mathbf{o}}_{l}\,,
\end{aligned}
\end{equation}
where $\mathbb{P}=\begin{bmatrix}
\cos\phi_{p} & -\sin\phi_{p}\\
\sin\phi_{p}&\cos\phi_{p}
\end{bmatrix}$ is the rotation matrix by angle $\phi_{p}$, which describes the phase gained from PCM. In order to achieve an acceleration measurement, we set the angle to $\phi_{p}=2n\pi$. $\mathbb{T}_{\rm{sag}}^{l,r}$ and $\mathbb{T}_{\rm{sag}}^{b.a.}$ are the transfer matrix and back-action only matrix of the Sagnac interferometer, 
\begin{equation}
\begin{aligned}
\mathbb{T}_{\rm{sag}}^{l,r}=e^{2i\beta_{\rm{sag}}}
\begin{bmatrix}
1     &  0\\
-\mathcal{K}_{\rm{sag}}/2     & 1
\end{bmatrix}\,,
\\
\mathbb{T}_{\rm{sag}}^{b.a.}=e^{2i\beta_{\rm{sag}}}
\begin{bmatrix}
0     &  0\\
-\mathcal{K}_{\rm{sag}}/2     & 0
\end{bmatrix}\,,
\end{aligned}
\end{equation}
where $\mathcal{K}_{\rm{sag}}=\frac{4\Theta\gamma\sin^{2}\beta}{\Omega^{2}(\gamma^{2}+\Omega^{2})}$ is the Kimble optomechanical factor for Sagnac interferometer\,\cite{2018LSA.....7...11D,kimble2001}. In $\mathcal{K}_{\rm{sag}}$,  $\Theta=\frac{8\omega P_{c}}{McL}$, where $\omega$ is the laser frequency and $P_{c}$ is the circulating power in each arm, $\beta=\arctan\frac{\Omega}{\gamma}$ is the phase gained by the light reflected from a single cavity, and $\beta_{sag}=2\beta+\pi/2$ is the phase gained of light reflecting from Sagnac interferometer with 
arm cavities. $\gamma=\frac{cT}{4L}$ is the half-bandwidth of the arm cavity, where $L$ is arm cavity length and $T$ is the power transitivity of input test mass. 
$\hat{\mathbf{R}}_{l,r}=e^{i\beta_{\rm{sag}}}\sqrt{\mathcal{K}_{\rm{sag}}}
\{0,1\}^{T}$ are the response vectors for the Sagnac interferometer. By solving the equations above, we can derive the full I/O-relation for the acceleration-meter,
\begin{equation}
\hat{\mathbf{o}}=\hat{\mathbf{o}}_{r}=e^{2i\beta_{\rm{acc}}}\begin{bmatrix}
1&0\\-\mathcal{K}_{\rm{acc}}&1
\end{bmatrix}\hat{\mathbf{i}}+e^{i\beta_{\rm{acc}}}\begin{bmatrix}
0\\\sqrt{2\mathcal{K}_{\rm{acc}}}
\end{bmatrix}\frac{x}{x_{SQL}}\,,
\end{equation}
where $\beta_{\rm{acc}}=4\beta$ and $\mathcal{K}_{\rm{acc}}=4\Omega^{2}\gamma^{2}/(\Omega^{2}+\gamma^{2})^{2}\mathcal{K}_{\rm{sag}}$.
With BHD\,\cite{2015PhRvD..92g2009S, 2004CQGra..21S1067F}, we can read an arbitrary quadrature of field $\mathbf{\hat{o}}$, given by $\mathbf{H}^{\rm{T}}\mathbf{\hat{o}}$. Here $\mathbf{H}=\{\cos\zeta,\sin\zeta\}^{\rm{T}}$, where $\zeta$ is the readout angle. The spectral density of the quantum noise of the whole interferometer can be calculated as,
\begin{equation}\label{sensitivity}
S=x_{\rm SQL}^{2}\frac{\mathbf{H}^{\dagger}\cdot\mathbb{T}\cdot\mathbb{S}^{in}\cdot\mathbb{T}^{\dagger}\cdot\mathbf{H}}{|\mathbf{H}\cdot\hat{\mathbf{R}}|^{2}}
\end{equation}
where $\mathbb{S}^{in}$ is the 2×2 spectral density matrix of the input field. The input field $\mathbb{S}^{in}$ can be written as:
\begin{equation}
\mathbb{S}^{in}=\begin{bmatrix}
e^{2r}&0\\0&e^{-2r}
\end{bmatrix}
\end{equation}
where $r$ is the frequency independent squeezing strength. When $r=0$, the light is in vacuum state. 
Finally, we can obtain the quantum noise power spectral density of the whole interferometer as
\begin{equation}
S=\frac{x_{SQL}^{2}}{2}\frac{e^{-2r}+[\cot\zeta-\mathcal{K}_{\rm{acc}}]^{2}e^{2r}}{\mathcal{K}_{\rm{acc}}}\,.
\end{equation}
When $\zeta$ satisfies $\cot\zeta=\mathcal{K}_{\rm{acc}}$, the spectral density is minimized.  In the ideal case, a technique which can realise frequency dependent readout angle is called variational readout\,\cite{kimble2001,yanbeiQND}.

\subsection{Quantum noise of acceleration-meter versus speed-meter and position-meter}\label{Quantum noise}
We demonstrate the example with the parameters of advanced LIGO\,\cite{2015} but without the signal recycling cavity. Also, the bandwidth is optimized for the acceleration-meter. The main parameters are as following: laser wavelength 1064\,nm, $L=4\,\rm{km}$, $M=40\,\rm{kg}$, bandwidth $\gamma=100*2\pi\,\rm Hz$, and circulating power $P_{c}=800\,\rm{kW}$. Additionally, there is no squeezing injected.

\begin{figure}
    \centering
    \includegraphics[width=0.7\columnwidth]{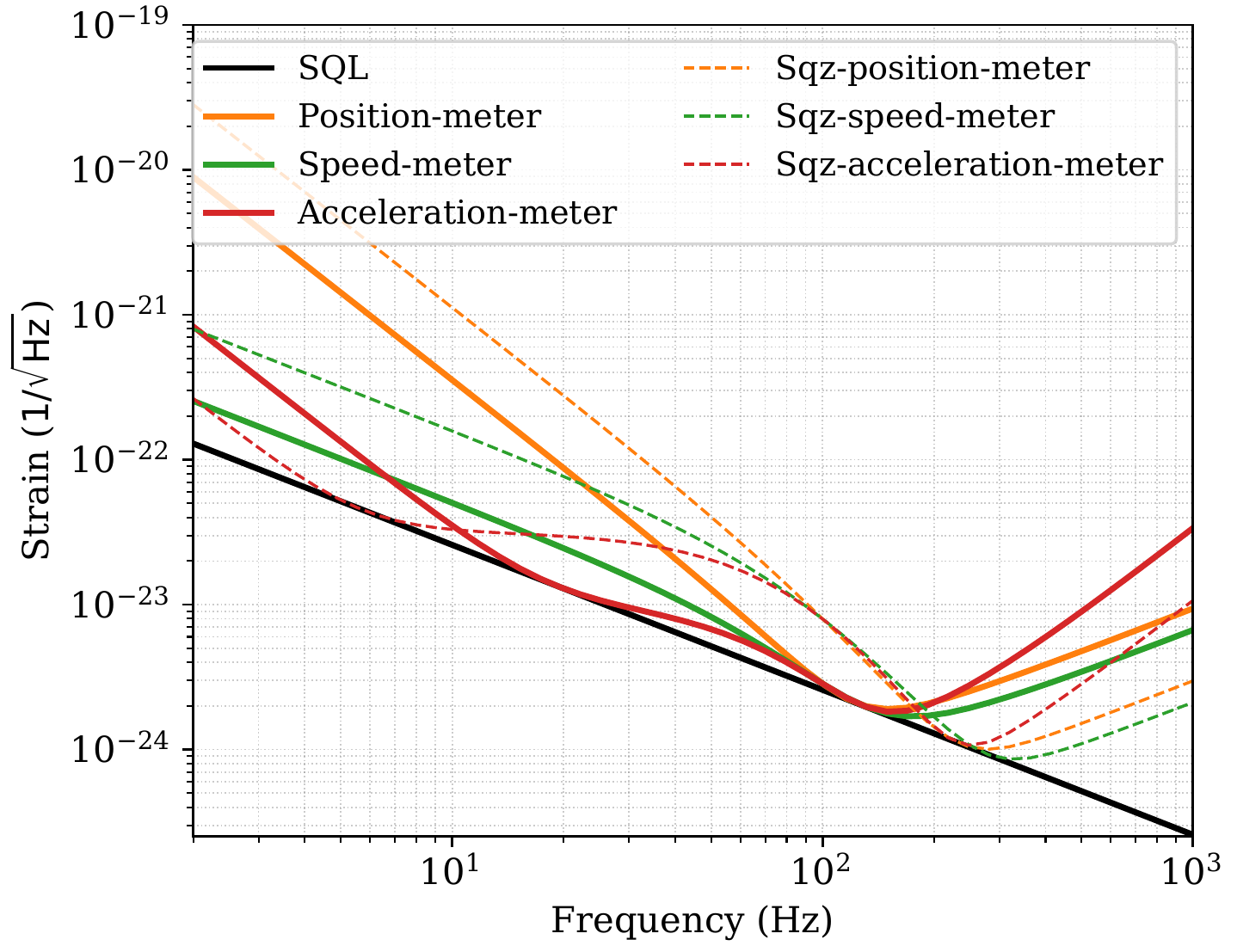}
    \caption{Comparison of quantum noise limited sensitivities(QNLS) of three types of interferometers: Michelson position-meter, Sagnac speed-meter and Polarisation-Sagnac acceleration-meter. They share common parameters, laser wavelength 1064\,nm, $L=4\,\rm{km}$, $M=40\,\rm{kg}$, bandwidth $\gamma=100\times 2\pi\, \rm Hz$ and circulating power $P_{c}=800\,\rm{kW}$. In this case, we set the homodyne angle such that $\cot\zeta=0$. Solid lines stand for the case with vacuum injected into the dark port. Dash lines stand for the case with squeezed vacuum (with squeeze factor $e^{-2r}=0.1$)injected into the dark port.
    }
    \label{fig:psd}
\end{figure}
\begin{figure}
    \centering
    \includegraphics[width=0.7\columnwidth]{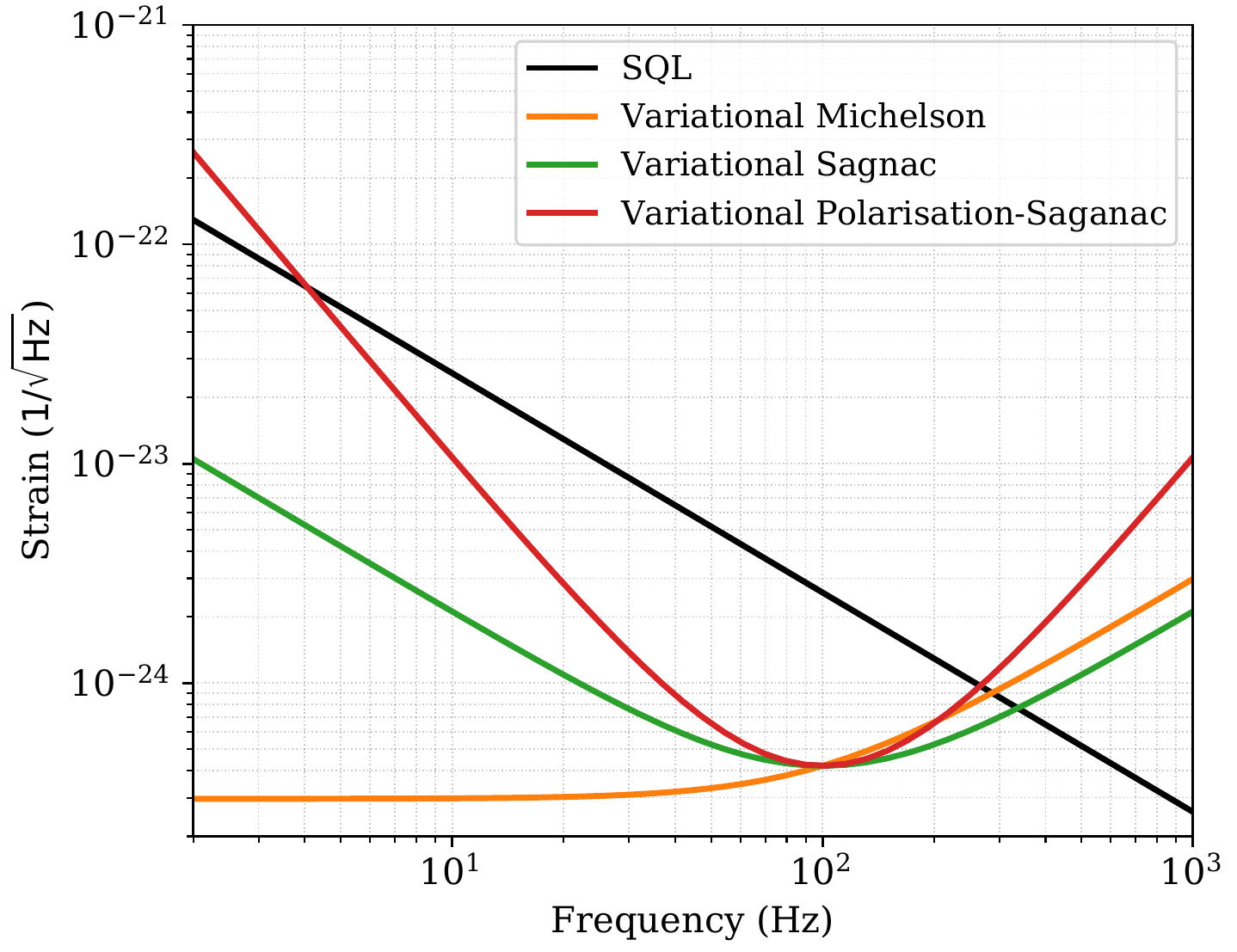}
    \caption{Comparison of quantum noise limited sensitivities of three types of interferometers including ideal variational readout, in which optimal readout angle is adopted.}
    \label{fig:varpsd} 
\end{figure}

In Fig.~\ref{fig:psd}, we plot the quantum noise limited sensitivity of three types of meters:  Michelson position-meter, Sagnac speed-meter and Polarisation-Sagnac acceleration-meter with $\cot\zeta=0$. Solid lines stand for the case with vacuum injected into the dark port. Dash lines stand for the case with 10dB phase quadrature squeezed vacuum
($e^{-2r}=0.1$)injected into the dark port. Fig.~\ref{fig:varpsd} shows the sensitivities when the readout angle can change along with Kimble optomechanical factor as $\cot\zeta=\mathcal{K}_{\rm acc}$. To simplify the calculation, we directly use the arm circulating power instead of the input power. It should be noted that to ensure they have same circulating power, the input power of Michelson position-meter should be set to twice that of the speed-meter and acceleration-meter. 

Instead of radiation pressure noise, the sensitivity of acceleration-meter is limited by shot noise at low frequencies (below 20\,Hz). The frequency dependence of the shot noise follows $1/\Omega^2$, which comes from the response of the acceleration to the equivalent displacement and is indicated by the first term of Eq.(\ref{SSacc}). The radiation pressure noise dominates between 20 and 100\,Hz. In the acceleration-meter, there are two frequencies, $\sim 20$ and $\sim 100$\,Hz, where the radiation pressure noise and shot noise are equal. 
The radiation pressure noise over $20$ to $50$\,Hz, which is constant against frequency is actually a feature of acceleration measurement following the second term of Eq.(\ref{SSacc}). With constant phase quadrature squeezing, the shot noise will be suppressed, while the radiation pressure noise will be amplified as shown by the squeezed position-and speed-meters in Fig.~\ref{fig:psd}. In the acceleration-meter, shot noise dominants both at low and high frequency and get suppressed by squeezing. At middle frequencies, the sensitivity becomes worse due to higher radiation pressure noises.
As shown in Fig.~\ref{fig:varpsd}, the radiation pressure noise can be evaded with ideal variational readout. However, with ideal variational readout, where only the shot noise couples to the measurement over the whole frequency band, the best scenario is position-meter which has the strongest response to the signal. 

\section{Conclusion}
In this article, we study the acceleration measurement for a gravitational wave detector and develop a new type of polarisation-Sagnac acceleration-meter interferometer that is based on the polarization circulation speed-meter and Sagnac speed-meter. We theoretically analyzed the noise of a lossless conceptual scheme. The main purpose of this configuration is to study stronger radiation pressure noise reduction in the low frequency range and add acceleration measurement as a possibility in the field of gravitational wave detection. Our analysis shows that the acceleration-meter can decrease the radiation pressure noise remarkably, but inevitably raises the shot noise. We have investigated the acceleration-meter in the context of gravitational wave detection for the first time and have established the possibility of broader research in the future.

\section{Acknowledgements}
We would like to thank Haixing Miao, Joe Bentley, Junxin Chen, Stefan Danilishin, Stefan Hild for fruitful discussions. T. Z. acknowledges the support of the Institute for Gravitational Wave Astronomy at University of Birmingham. J.L. acknowledges the support from the Department of Astronomy at Beijing Normal University. 

\appendix

\section{Position measurement}\label{Position measurement}
The main purpose of this appendix is to derive the power spectrum of the quantum noise of the position-meter. 
In the position measurement, shown in left part of Fig.~\ref{fig:accelerationmeter}, the phase of the light is modulated by a free test mass when the light is reflected by one side of test mass. We can write the phase detected as:
\begin{equation}
\phi(t) = \phi_{0}(t)+2kx(t)\,,
\end{equation}
where $\phi_{0}$ is the detected phase when the test mass is static. $k=\omega/c$, is the wave number, and $x$ is the displacement of the test mass. With a homodyne detection, we can access all quadratures of the light. Similar to in\,\cite{2012LRR....15....5D}, the readout signal in frequency domain can be described by the following equation,
\begin{equation}\label{homodyne readout}
\hat{o}(\Omega)=-\phi(\Omega)\sin\zeta+\frac{\hat{\mathcal{I}}(\Omega)-\mathcal{I}_{0}}{2\mathcal{I}_{0}}\cos\zeta\,,
\end{equation}
where 
$\hat{\mathcal{I}}(\Omega)$ is the light intensity and $\mathcal{I}_{0}$ is the average light intensity.
The displacement extracted from the readout signal can be written as below, consisting of a signal part and fluctuation part:
\begin{equation}
x(\Omega)= -\frac{\hat{o}(\Omega)}{2k\sin\zeta}=x_{\rm{sig}}(\Omega)+x_{\rm{flu}}(\Omega)\,,
\end{equation}
where
\begin{equation}
x_{\rm flu}(\Omega)=\frac{1}{2k}[\phi_{0}(\Omega)-\frac{\hat{\mathcal{I}}(\Omega)-\mathcal{I}_{0}}{2\mathcal{I}_{0}}\cot\zeta]\,.
\end{equation}
The spectral density of the measurement noise can be then derived as
\begin{equation}
S_{x}(\Omega)=\frac{1}{4k^{2}}(S_{\phi}+\frac{S_{\mathcal{I}}}{4\mathcal{I}^{2}_{0}}cot^{2}\zeta)\,,
\end{equation}
where $S_{\phi}$ and $S_{\mathcal{I}}$ are the spectral density of the phase and intensity of the light field. 
The back action force of the light onto the test mass can be written as
\begin{equation}
\hat{F}(\Omega)=\frac{2(\hat{\mathcal{I}}(\Omega)-\mathcal{I}_{0})}{c}\,.
\end{equation}
The resulting spectral density of the back action and cross-correlation noise are
\begin{equation}
\begin{aligned}
S_{F}(\Omega)=\frac{4S_{\mathcal{I}}(\Omega)}{c^{2}}\,,\\
S_{xF}(\Omega)=\frac{S_{\mathcal{I}}(\Omega)}{2\omega\mathcal{I}_{0}}cot\zeta\,.
\end{aligned}
\end{equation}

According to the above equations, we can find that $S_{x}$ and $S_{F}$ follows the Schr\"{o}dinger–Robertson  uncertainty relation:
\begin{equation}
S_{x}(\Omega)S_{F}(\Omega)-{S_{xF}}^{2}(\Omega)=\frac{S_{\phi}(\Omega)S_{\mathcal{I}}(\Omega)}{\omega^{2}}\geq\frac{\hbar^{2}}{4}\,.
\end{equation}

If there is no correlation between the back-action and measurement fluctuations, i.e., $S_{xF}(\Omega)=0$, and we apply the uncertain relation to Eq.(\ref{sd}), we obtain:
\begin{equation}
S(\Omega)\geq S_{x}+|\chi(\Omega)|^2\frac{\hbar^{2}}{4S_{x}}\geq\frac{\hbar}{M\Omega^{2}}\,.
\end{equation}
When $S_{xF}\neq0$, which could be achieved by tuning readout angle through variational readout\,\cite{braginsky_khalili_thorne_1992}, we can overcome the SQL.

\section{Acceleration measurement}\label{Acceleration measurement}
Differ to position measurement, in speed measurement, the light reflects on both side and is modulated twice by the test mass. Consequently, speed measurement can be conducted by measuring the position of the test masses in two sequential moments. Similarly, the acceleration measurement can be conducted by measuring speed of the test masses at two moments. 
As the conceptual scheme shows in the right part of Fig.~\ref{fig:accelerationmeter}, the light hits the test mass 4 times and is phase modulated by the test mass motion. The process can be described by
\begin{equation}\label{xacc}
\begin{aligned}
\phi(t) =& \phi_{0}(t)+2k[x(t)-x(t-\tau)]\\
&-2k[x(t-\tau)-x(t-2\tau)] \\
=&\phi_{0}(t)+2k\tau v(t)-2k\tau v(t-\tau)\\
=&\phi_{0}(t)+2k\tau^{2}a(t)\,,
\end{aligned}
\end{equation}
while
\begin{equation}
\begin{aligned}
v(t)&=\frac{x(t)-x(t-\tau)}{\tau}\\
a(t)&=\frac{v(t)-v(t-\tau)}{\tau}\,,
\end{aligned}
\end{equation}
where $\tau$ is the time delay between reflections as shown in Fig.~\ref{fig:accelerationmeter}, $v$ is the speed and $a$ is the acceleration.

The light traveling back to the phase detector gains a phase shift proportional to the difference of two velocities with interval time $\tau$, which results in an acceleration measurement of the motion of the test mass. With a homodyne detection, substituting Eq.\,\ref{xacc} into Eq.\,\ref{homodyne readout}, and the displacement can be written in spectral representation as:
\begin{equation}
x^{\rm acc}=-\frac{\hat{o}(\Omega)}{2k\sin\zeta(1-e^{-i\Omega\tau})^{2}}=x^{\rm acc}_{\rm sig}(\Omega)+x^{\rm acc}_{\rm flu}(\Omega)\,,
\label{a}
\end{equation}
where
\begin{equation}\label{xaccflu}
x^{\rm acc}_{\rm flu}(\Omega)=\frac{x_{\rm flu}}{(1-e^{-i\Omega\tau})^{2}}\,.
\end{equation}
The $(1-e^{-i\Omega\tau})^{2}$ in the denominator comes from the 4 reflections as mentioned before. Since the gravitational wave signal period is much longer than $\tau$ here, we have $\Omega\tau\ll1$. Therefore, Eq.(\ref{xaccflu}) can be approximately written as: 
\begin{equation}
x^{\rm acc}_{\rm flu}(\Omega)\approx\frac{-x_{\rm flu}(\Omega)}{\Omega^{2}\tau^{2}}\,.
\label{b}
\end{equation}
The back-action force for the light reflected four times onto the test mass can be written as:
\begin{equation}
\hat{F}^{\rm acc}(t)=\frac{2}{c}[\hat{\mathcal{I}}(t+2\tau)-\hat{\mathcal{I}}(t+\tau)]-[\hat{\mathcal{I}}(t+\tau)-\hat{\mathcal{I}}(t)]\,.
\end{equation}
In spectral representation, it can be written as:
\begin{equation}
\hat{F}^{\rm acc}(\Omega)=\frac{2}{c}(e^{i\Omega\tau}-1)^{2}\hat{\mathcal{I}}=(e^{i\Omega\tau}-1)^{2}\hat{F}(\Omega)\,.
\end{equation}
Similar to $x^{\rm acc}_{\rm flu}$, $\hat{F}^{\rm acc}$ can be approximately written as:
\begin{equation}\label{eq:Facc}
\hat{F}^{\rm acc}(\Omega)\approx -\tau^{2}\Omega^{2}{F}(\Omega)\,.
\end{equation}
The relation of the spectral densities in the acceleration measurement and position measurement can then be calculated as below:
\begin{equation}
\begin{aligned}\label{Sacc}
S_{x}^{\rm acc}(\Omega)&=\frac{S_{x}(\Omega)}{\Omega^{4}\tau^{4}}\,, \\
S_{F}^{\rm acc}(\Omega)&=\tau^{4}\Omega^{4}S_{F}(\Omega)\,, \\
S_{xF}^{\rm acc}(\Omega)&=S_{xF}(\Omega)\,.
\end{aligned}
\end{equation}
Substituting Eq.\,(\ref{Sacc}) into Eq.\,\ref{sd}, we get
\begin{equation}
S_{\rm acc}(\Omega)=\frac{S_{x}}{\Omega^{4}\tau^4}+\Omega^4\tau^{4}|\chi(\Omega)|^{2}S_{F}-2\chi(\Omega)S_{xF}\,.
\end{equation}
In the case of $S_{xF}^{acc}=0$, we find that:
\begin{equation}
S_{\rm acc} \geq \frac{\hbar}{M\Omega^2}\,,
\end{equation}
which denotes the SQL in displacement of free mass.

\bibliography{bibliography}
\end{document}